\documentclass[]{spie}  

\usepackage{amsmath,amsfonts,amssymb}
\usepackage{graphicx}
\usepackage[colorlinks=true, allcolors=blue]{hyperref}

\title{The Robo-AO KOI Survey: laser adaptive optics imaging of every Kepler exoplanet candidate}

\author[a]{Carl Ziegler}
\author[a]{Nicholas M. Law}
\author[b]{Christoph Baranec}
\author[c]{Tim Morton}
\author[d]{Reed Riddle}
\author[b]{Dani Atkinson}
\author[b]{Larissa Nofi}

\affil[a]{Department of Physics and Astronomy, University of North Carolina at Chapel Hill, Chapel Hill, NC 27599-3255, USA}
\affil[b]{Institute for Astronomy, University of Hawai`i at M\={a}noa, Hilo, HI 96720-2700, USA}
\affil[c]{Department of Astrophysical Sciences, Princeton University, Princeton, NJ 08544, USA}
\affil[d]{Division of Physics, Mathematics, and Astronomy, California Institute of Technology, Pasadena, CA 91125, USA}

\authorinfo{Send correspondence to carlziegler@unc.edu}

\begin{document} 
\maketitle

\begin{abstract}
The Robo-AO \textit{Kepler} Planetary Candidate Survey is observing every \textit{Kepler} planet candidate host star (KOI) with laser adaptive optics imaging to hunt for blended nearby stars which may be physically associated companions. With the unparalleled efficiency provided by the first fully robotic adaptive optics system, we perform the critical search for nearby stars (0.15" to 4.0" separation with contrasts up to 6 magnitudes) that dilute the observed planetary transit signal, contributing to inaccurate planetary characteristics or astrophysical false positives. We present 3313 high resolution observations of \textit{Kepler} planetary hosts from 2012-2015, discovering 479 nearby stars. We measure an overall nearby star probability rate of 14.5$\pm$0.8$\%$. With this large data set, we are uniquely able to explore broad correlations between multiple star systems and the properties of the planets which they host, providing insight into the formation and evolution of planetary systems in our galaxy. Several KOIs of particular interest will be discussed, including possible quadruple star systems hosting planets and updated properties for possible rocky planets orbiting with in their star's habitable zone. 
\end{abstract}

\keywords{binaries: close \-- instrumentation: adaptive optics \-- techniques: high angular resolution \-- methods: data analysis \-- methods: observational}

\section{INTRODUCTION}
\label{sec:intro}

Using precision photometry, the \textit{Kepler} mission discovered approximately 4700 planetary candidates. By detecting the periodic dips in the host stars brightness consistent with a planet transit, the candidate exoplanets require follow-up observations to rule out astrophysical false positives and for host star characterization.

With relatively coarse resolution (pixel size of $\sim$4") \cite{haas10}, the photometric apertures of \textit{Kepler} are frequently polluted by nearby stars, either associated or unassociated foreground or background stars.  The majority of solar-type stars do form with at least one companion star \cite{duquennoy91, raghavan10}, and these nearby stars can result in false-positive transiting events. In the case of a bona fide exoplanet, the diluted transit signal results in inaccurate estimates of planetary characteristics.  High-angular-resolution imaging follow-ups are therefore a crucial step in the exoplanet validation process to distinguish multiple stellar systems and subsequently confirm or reject planetary candidates.

There has been considerable effort by the community to perform high-resolution imaging surveys of the KOIs \cite{howell11, adams12, adams13, lillo12, lillo14, horch12, marcy14, dressing14, wang15a, wang15b, torres15, everett15, kraus16}.  These surveys, however, have been over a relatively small number of targets compared to the full set of \textit{Kepler} planetary candidates.  This piecemeal approach leads to inconsistent vetting, while limiting the comprehensive statistics and correlations that can be derived from a large dataset of high resolution images of multiple stellar systems hosting planetary systems.

A complete, consistent high-resolution survey of all the the KOIs with ground-based adaptive optics (AO) is limited by the typical overheads required with traditional systems.  Utilizing the order-of-magnitude increase in time-efficiency provided by Robo-AO, the first robotic laser adaptive optics system, we have performed high-resolution imaging of every KOI system.  The first paper in this survey\cite[hereafter Paper I]{law14} observed 715 \textit{Kepler} planetary candidates, identifying 53 companions\footnote{For brevity, we denote stars which we found within our detection radius of KOIs as ``companions,'' in the sense that they are asterisms associated on the sky.}, with 43 new discoveries, for a binary fraction rate of 7.4\%$\pm$1.0\% within separations of 0.15" to 2.5".  The second paper in this survey\cite[hereafter Paper II]{baranec16} observed 969 \textit{Kepler} planetary candidates, identifying 202 companions, with 139 new discoveries, for a binary fraction rate of 11.0\%$\pm$1.1\% within separations of 0.15" to 2.5", and 18.1\%$\pm$1.3\% within separations of 0.15" to 4.0".  The third paper in the survey\cite[hereafter Paper III]{ziegler16} looked at a total of 1629 targets observed, finding 223 companions around 206 KOIs, 209 of which have not been previously imaged in high resolution, for a multiplicity fraction of 12.6\%$\pm$0.9$\%$ within 4.0" of planetary candidate hosting stars.

\begin{figure*}
\centering
\includegraphics[width=0.53\paperwidth]{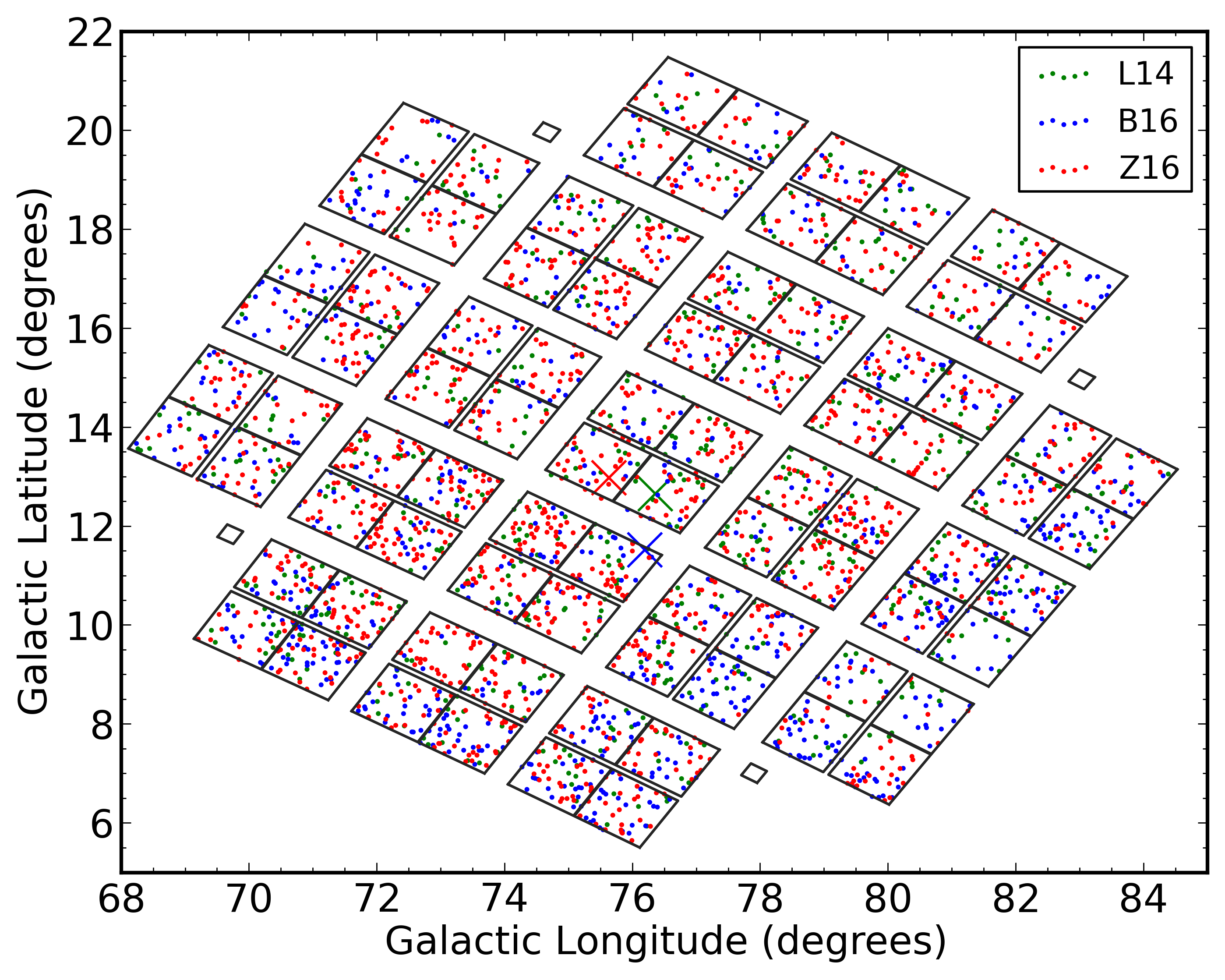}
\caption{Location on sky of observed KOIs from Paper I (L14)\cite{law14}, Paper II (B16)\cite{baranec16}, and Paper III (Z16)\cite{ziegler16}.  The median coordinates of the targeted KOIs is designated by an `$\times$'.  A projection of the \textit{Kepler} field of view is provided for reference.}
\label{fig:fov}
\end{figure*}

We begin in Section \ref{sec:observations} by describing our target selection, the Robo-AO system, and follow-up observations. In Section \ref{sec:datareduction} we describe the Robo-AO data reduction and the companion detection and analysis.  In Section \ref{sec:Discoveries} we describe the results of this survey, including discovered companions, and discuss the results, detailing the effects on the planetary characteristics of the survey's discoveries and looking at the overall binarity statistics of the \textit{Kepler} planet candidates.  We conclude in Section \ref{sec:conclusion}.

\section{OBSERVATIONS}
\label{sec:observations}

\begin{figure*}
\centering
\includegraphics[width=0.83\paperwidth]{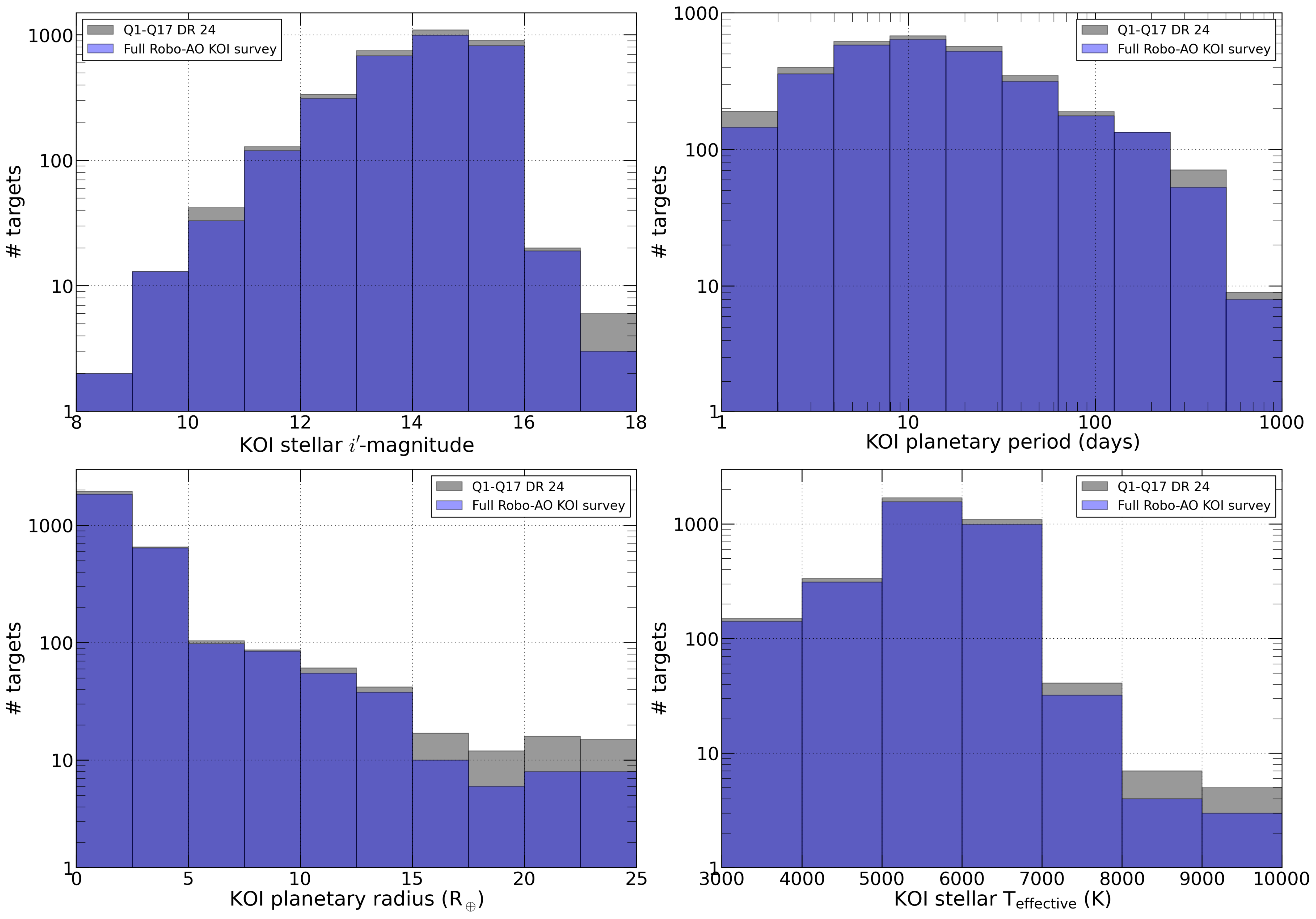}
\caption{Comparison of the distribution of the combined Robo-AO survey (Paper I, Paper II, and Paper III) to the complete set of KOIs from Q1-Q17 \cite{borucki10, borucki11a, borucki11b, batalha13, burke14, rowe14, coughlin15}.}
\label{fig:histograms}
\end{figure*}

KOI targets were selected from the KOI catalog based on Q1-Q17 \textit{Kepler} data\cite{borucki10, borucki11a, borucki11b, batalha13, burke14, rowe14, coughlin15}.  KOIs flagged as false positives using \textit{Kepler} data were removed.  In Figure$~\ref{fig:histograms}$, the properties of the targeted KOIs for the full Robo-AO survey as of date are compared to the set of all KOIs from Q1-Q17 with CANDIDATE dispositions based on \textit{Kepler} data.  On-sky locations of targeted KOIs are plotted in Figure$~\ref{fig:fov}$.

The first robotic laser guide star adaptive optics system, the automatic Robo-AO system\cite{baranec13, baranec14} can efficiently perform large, high-resolution surveys. The AO system runs at a loop rate of 1.2 kHz to correct high-order wavefront aberrations, delivering a median Strehl ratio of 9\% in the \textit{i}\textsuperscript{$\prime$}-band.  Observations were taken in either a \textit{i}\textsuperscript{$\prime$}-band filter or a long-pass filter cutting on at 600 nm (LP600 hereafter).  The LP600 filter approximates the \textit{Kepler} passband at redder wavelengths, while also suppressing blue wavelengths that reduce adaptive optics performance.

Typical seeing at the Palomar Observatory is between 0.8" and 1.8", with median around 1.1" \cite{baranec14}. The typical FWHM (diffraction limited) resolution of the Robo-AO system is 0.15". Images are recorded on an electron-multiplying CCD (EMCCD), allowing short frame rates for tip and tilt correction in software using a natural guide star ($m_V < 16$) in the field of view.  Specifications of the Robo-AO KOI survey are summarized in Table$~\ref{tab:specs}$.

\begin{table}[ht]
\caption{\label{tab:survey_specs}The specifications of the Robo-AO KOI survey}
\begin{center}
\begin{tabular}{ll}
\hline
KOI targets    	& 3313 \\
Observation wavelengths & 600-950 nm \\
FWHM resolution   	& 0.15" \\
Field size & 44" $\times$ 44"\\
Detector format       	& 1024$^2$ pixels\\
Pixel scale & 43.1 mas / pix\\
Exposure time & 90 seconds \\
Targets observed / hour & 20\\
Observation dates & 2012 June 17 --\\ &  2015 June 12\\
\hline
\label{tab:specs}
\end{tabular}
\end{center}
\end{table}

\section{DATA REDUCTION}
\label{sec:datareduction}

With the largest adaptive optics dataset yet assembled by Robo-AO, the data reduction process was automated as much as possible for efficiency and consistency. After initial pipeline reductions, the target stars were identified and cutouts prepared as described in Section \ref{sec:pipeline}, companions automatically identified (Section \ref{sec:compsearch}), PSF subtraction performed and companions again auto-identified (Section \ref{sec:psfsubtraction}), and constraints measured on the companion sensitivity of the survey (Section \ref{sec:imageperf}).  Finally, the properties of the detected companions are measured in Section \ref{sec:characterization}.

\subsection{Imaging Preparation}
\label{sec:pipeline}

The Robo-AO imaging pipeline \cite{law09, law14} reduced the images: the raw EMCCD output frames are dark-subtracted and flat-fielded and then stacked and aligned using the Drizzle algorithm \cite{fruchter02}, which also up-samples the images by a factor of two.  To avoid tip/tilt anisoplanatasism effects, the image motion was corrected by using the KOI itself as the guide star in each observation.

To verify that the star viewed in the image is the desired KOI target, we created Digital Sky Survey cutouts of similar angular size around the target coordinates.  Each image was manually checked to assure no ambiguity in the target star with images with either poor performance or incorrect fields removed.  These bad images made up approximately 2$\%$ of all our images, and for all but 2 of the targets additional images were available. 

To facilitate the automation of the data reduction, centered 8.5" square cutouts were created around the 1629 verified target KOIs.  We select a 4" separation cutoff for our companion search to detect all nearby stars that would blend with the target KOI in a \textit{Kepler} pixel.

\begin{figure}
\centering
\includegraphics[width=445pt]{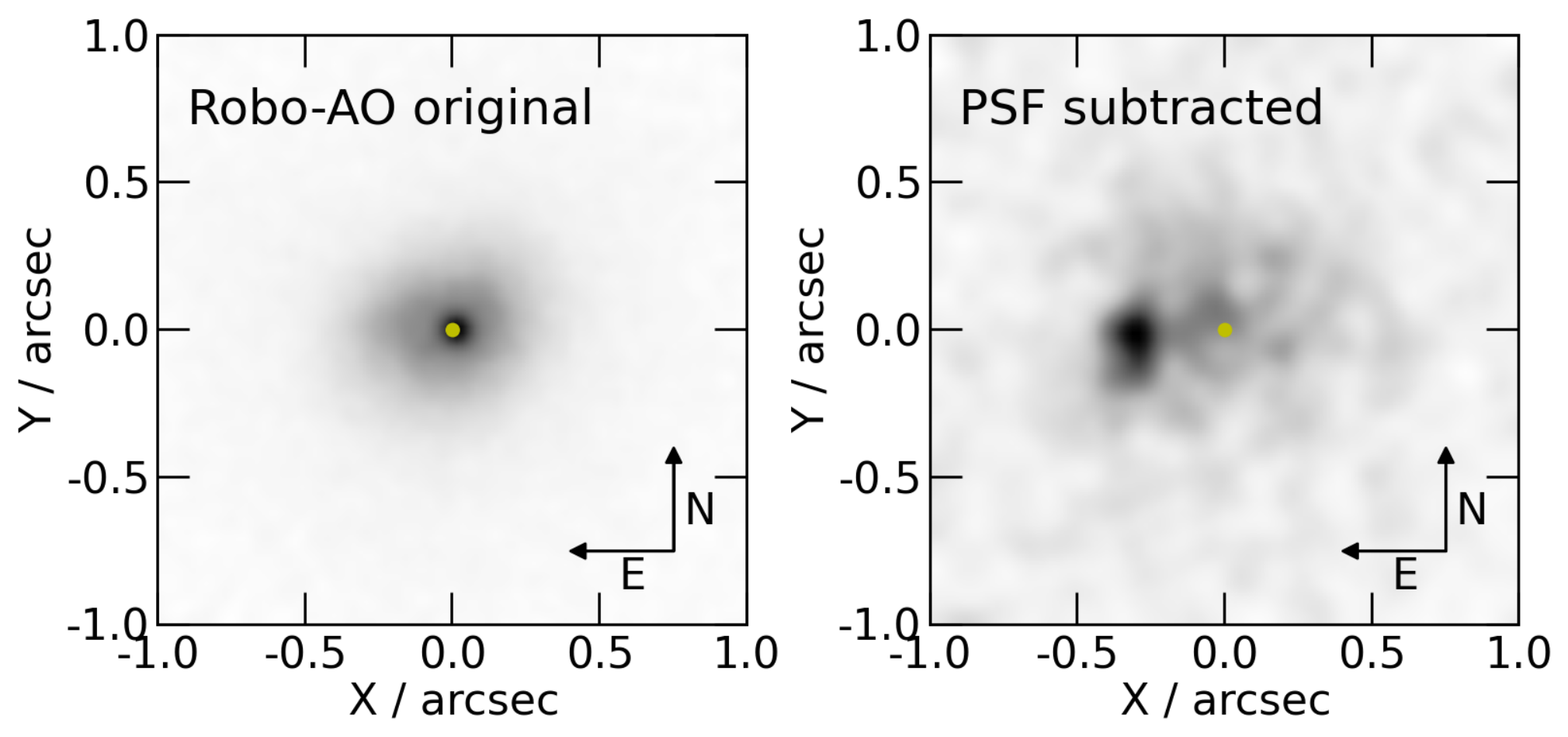}
\caption{Example of PSF subtraction on KOI-5762 with companion separation of 0.34".  The yellow circle marks the position of the primary star's PSF peak. Both images have been scaled and smoothed for clarity.  Successful removal of the PSF leaves residuals consistent with photon noise.  The 2" square field shown here is approximately equal to half the \textit{Kepler} pixel size.}
\label{fig:psf}
\end{figure}

\subsection{PSF Subtraction}
\label{sec:psfsubtraction}

To identify close companions, a custom locally optimized point spread function (PSF) subtraction routine based on the Locally Optimized Combination of Images algorithm \cite{lafreniere07} was applied to centered cutouts of all stars.  The code uses a set of twenty KOI observations as reference PSFs, as it is improbable that a companion would appear at the same position in two different images.  A locally optimized PSF is generated and subtracted from the original image, leaving residuals consistent with photon noise.

This procedure was performed on all KOI images out to 2", and the results visually checked for companions.  Figure$~\ref{fig:psf}$ shows an example of the PSF subtraction performance.  The PSF subtracted images were subsequently ran through the automated companion finding routine, as described in Section \ref{sec:compsearch}.

\subsection{Companion Detection}
\label{sec:compsearch}

An initial visual companion search was performed redundantly by three of the authors.  This search yielded a preliminary companion list, and filtered out bad images.

After initial visual inspection and culling, we ran all images through a custom automated search algorithm, based on the code described in Paper I. The algorithm slides a 5-pixel diameter aperture within concentric annuli centered on the target star. Any aperture with $>$+5$\sigma$ outlier to the local noise is considered a potential astrophysical source. These are subsequently checked manually, eliminating spurious detections with dissimilar PSFs to the target star and those having characteristics of a cosmic ray hit, such as a single bright pixel or bright streak.

Multiple possible companions visually identified but fell beneath the formal 5$\sigma$ required for a discovery.  Despite not reaching our formal significance level required for a discovery, previous results suggest that all but a small fraction are likely real: Keck NIRC2 observations have confirmed all 15 `likely' detections in Paper I and 38 `likely' companions in Paper II.

\subsection{Imaging Performance Metrics}
\label{sec:imageperf}

The two dominant factors that effect the image performance of the Robo-AO system are seeing and target brightness.  An automated routine was used to classify the image performance for each target using PSF core size as a proxy for image performance, and binning the targets into three performance groups: low, medium and high.

\begin{figure*}
\centering
\includegraphics[width=0.75\paperwidth]{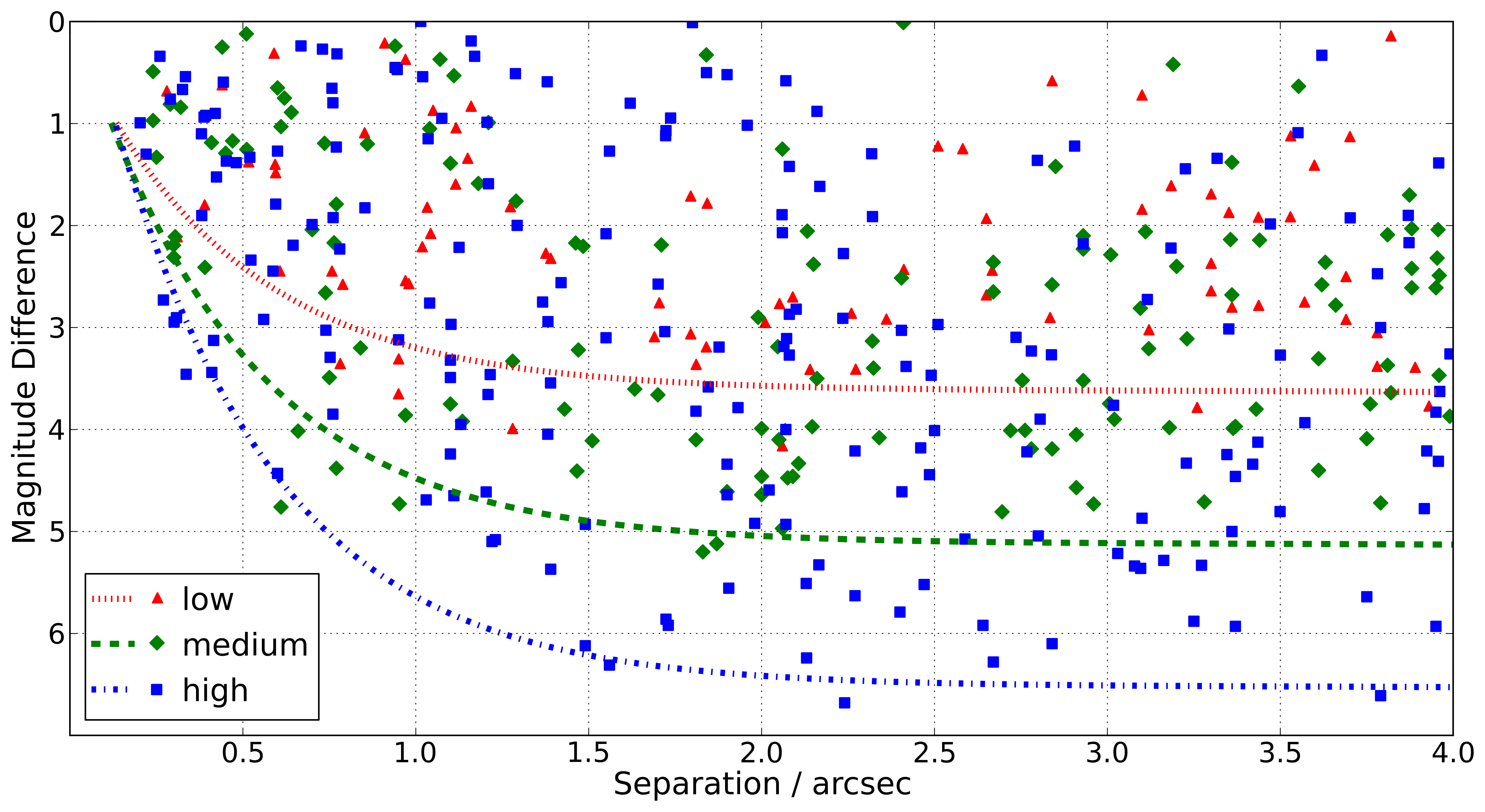}
\caption{Separations and magnitude differences of the detected companions in the full Robo-AO KOI survey, with the color and shape of each star denoting the associated typical low-, medium- and high-performance 5$\sigma$ contrast curve during the observation (as described in Section \ref{sec:imageperf}).}
\label{fig:contrastcurves}
\end{figure*}

We determine the angular separation and contrast consistent with a 5$\sigma$ detection by injecting artificial companions, a clone of the primary PSF.  For concentric annuli of 0.1" width, the detection limit is calculated by steadily dimming the artificial companion until the auto-companion detection algorithm (Section \ref{sec:compsearch}) fails to detect it.  This process is subsequently performed at multiple random azimuths within each annulus and the limiting 5$\sigma$ magnitudes are averaged. For clarity, these average magnitudes for all radii measurements are fitted with functions of the form $a*sinh(b*r+c)+d$ (where \textit{r} is the radius from the target star and \textit{a, b, c} and \textit{d} are fitting variables).  Contrast curves for the three performance groups are shown in Figure$~\ref{fig:contrastcurves}$.

\subsection{Companion Characterization}
\label{sec:characterization}

\subsubsection{Contrast Ratios}
\label{sec:contrastratios}

For wide, resolved companions with little PSF overlap, the companion to primary star contrast ratio was determined using aperture photometry on the original images. The aperture radius was cycled in one-pixel increments from 1-5 FWHM for each system, with background measured opposite the primary from the companion (except in the few cases where another object falls near or within this region in the image).  Photometric uncertainties are estimated from the standard deviation of the contrast ratios measured for the various aperture sizes.

For close companions, the estimated PSF was used to remove the blended contributions of each star before aperture photometry was performed.  The locally optimized PSF subtraction algorithm can attempt to remove the flux from companions using other reference PSFs with excess brightness in those areas.  For detection purposes, we use many PSF core sizes for optimization, and the algorithm's ability to remove the companion light is reduced. However, the companion is artificially faint as some flux has still been subtracted. To avoid this, the PSF fit was redone excluding a six-pixel-diameter region around the detected companion.  The large PSF regions allow the excess light from the primary star to be removed, while not reducing the brightness of the companion.

\subsubsection{Separation and Position Angles}
\label{sec:separationposangles}

Separation and position angles were determined from the raw pixel positions.  Uncertainties were found using estimated systematic errors due to blending between components.  Typical uncertainty in the position for each star was 1-2 pixels.  Position angles and the plate scale were calculated using a distortion solution produced using Robo-AO measurements for the globular cluster M15.\footnote{S. Hildebrandt (2013, private communication).}

\section{DISCOVERIES}
\label{sec:Discoveries}

We find 479 stars nearby 3313 \textit{Kepler} planetary candidates, yielding a nearby star fraction within 4" of 14.5$\pm$0.8$\%$.  Approximately half (48.7\%) of the nearby stars are within 2", a separation range where only high-resolution surveys are able to accurately measure the properties of the companion stars.  In addition, these near stars ($\rho<$2") also have the highest probability of being associated.  We find that  in Papers I, II, and III, respectively, 6.4$\%$, 8.2$\%$, 6.8$\%$ of KOIs have nearby stars within 2".  The disparity in these nearby star rates is likely to due to the disparity in the median location
of targeted KOIs, as plotted in Figure$~\ref{fig:fov}$.  Targets from Paper II lie closer to the Galactic plane than Papers I and III, an area with higher stellar densities and thus a higher probability of unassociated stars being found nearby the KOI.

\begin{figure}
\centering
\includegraphics[width=285pt]{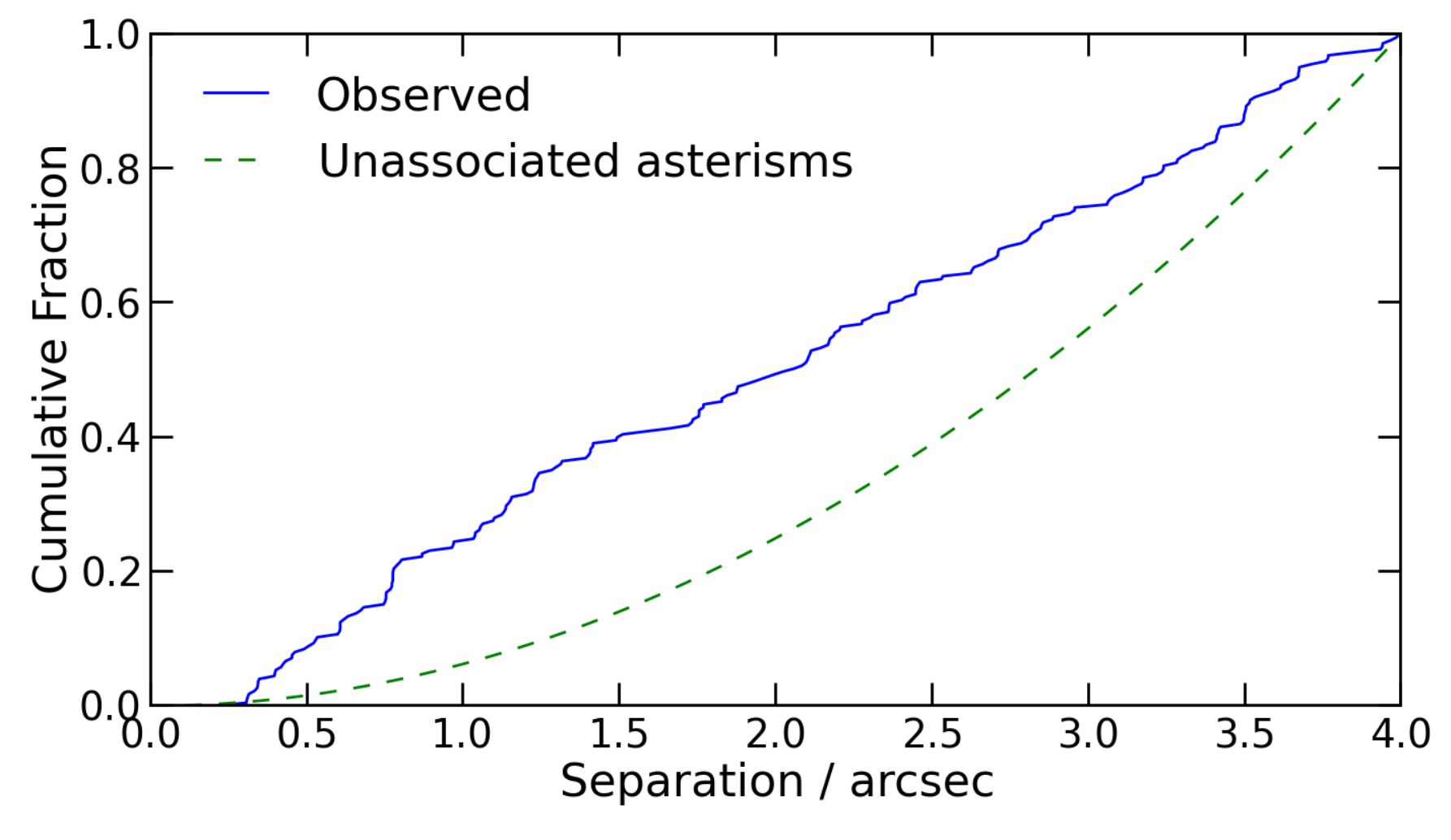}
\caption{The cumulative distribution of nearby stars within a given separation from our observations in Paper II and Paper III, and the expected distribution from a set of the same number of unassociated stars.  For all separations, the observed number of companions we detected is above the expected number if all stars were unassociated.}
\label{fig:sep_dist}
\end{figure}

A full probability of association analysis is forthcoming in subsequent papers in the survey, however we expect a small number of unassociated asterisms within our complete set of observed targets.  A recent follow-up study\cite{atkinson16} of 104 nearby stars to KOIs discovered by Robo-AO observed with NIRC2 on Keck II found that 6-8$\%$ of the candidate companions are physically unassociated at greater than 5$\sigma$ significance level. A statistical argument for the majority of nearby stars being associated is derived from the observed distribution of companion separations: if all companions were unassociated background or foreground stars, we would expect a quadratic distribution of companions (i.e., $\sim$4$\times$ the number of objects within 4" as at 2").  Instead we find a near linear distribution. The dissimilarity between the observed distribution and the distribution of all unassociated objects is shown in Figure$~\ref{fig:sep_dist}$. This survey also has less sensitivity at low-separations (see 5$\sigma$ contrast curves plotted in Figure$~\ref{fig:contrastcurves}$), so when correcting for completeness, the disparity from the expected distribution if all stars were unassociated asterisms will only increase. We therefore expect the overall multiplicity trends to remain relatively unchanged when the unassociated objects are removed.

\begin{figure*}
\centering
\includegraphics[width=400pt]{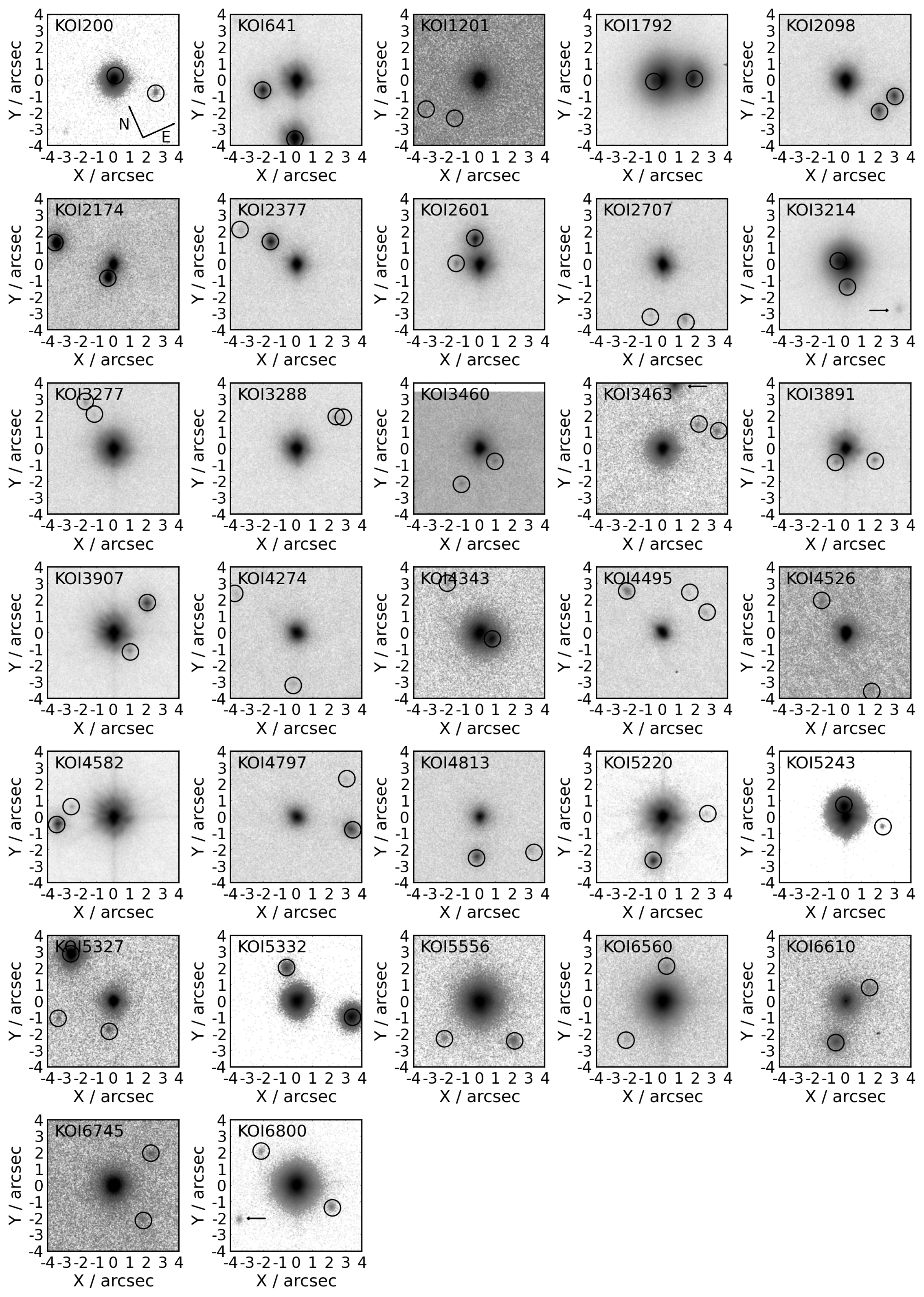}
\caption{Normalized log-scale cutouts of 32 KOIs with multiple companions with separations $<$4" resolved with Robo-AO.  The angular scale and orientation (displayed in the first frame) is similar for each cutout, and circles are centered on the detected nearby stars.  Three targets (KOIs 3214, 3463, and 6800) have a possible third companion, marked with arrows, outside our 4" separation cutoff, as described in Section$~\ref{sec:quadsystems}$.}
\label{fig:multiplesgrid}
\end{figure*}

\subsubsection{Possible Quadruple Systems}
\label{sec:quadsystems}

We find three nearby stars to KOIs 4495 and 5327.  If all stars are bound and the planet is confirmed, these systems would be the third and fourth known planets residing in quadruple star systems. One of the two previous confirmed planet-hosting quadruple systems was also first discovered with Robo-AO\cite{roberts15}. We also find three systems (KOIs 3214, 3463, and 6800) with two nearby stars within our survey search radius of 4", and an additional third star just outside our separation cutoff. The Robo-AO images of the four possible quadruple systems from this work are displayed in Figure$~\ref{fig:multiplesgrid}$

\subsection{Stellar Multiplicity and \textit{Kepler} Planet Candidates}

With this large, unprecedented dataset of hundreds of multiple star systems hosting planetary candidates, we are able to search for insight into the process of planetary formation and evolution.  A nearby star may disturb or disrupt formation of planets of various sizes or orbital periods through perturbations, or might lead to orbital migration or even ejection of formed planets as the system evolves.  The presence of a nearby star within the \textit{Kepler} aperture will also change the calculated planetary characteristics, altering our understanding of the layout of the system.

\subsubsection{Habitable Zone Candidates}

A major goal of the \textit{Kepler} mission is to find rocky exoplanets orbiting within the habitable zone of their host star, thus able to support life.  The presence or absence of stellar companions within an exoplanet system allows more precise estimates of the orbit and size of the exoplanets discovered.  The exact requirements for habitability are still debated \cite{kasting93, selis07, seager13, zsom13}, however it has been shown that the transition between ``rocky'' and ``non-rocky'' occurs rather sharply at R$_{P}$=1.6$R_{\oplus}$ \cite{rogers15}.   Overall, the existence of an unknown stellar companion within the same photometric aperture as the KOI will increase the calculated radius of the planet, as the observed transit signal will be diluted by the constant light of the nearby star.

In Paper II and Paper III, we detected companions to 23 KOIs which host planetary candidates with equilibrium temperatures, from the NEA, in the HZ range (273 K$\le$T$_{eq}$$\le$373 K) and R$<$4 $R_{\oplus}$.  All are newly detected in this survey.

\subsubsection{Stellar Multiplicity and KOI Number}

\begin{figure}
\centering
\includegraphics[width=285pt]{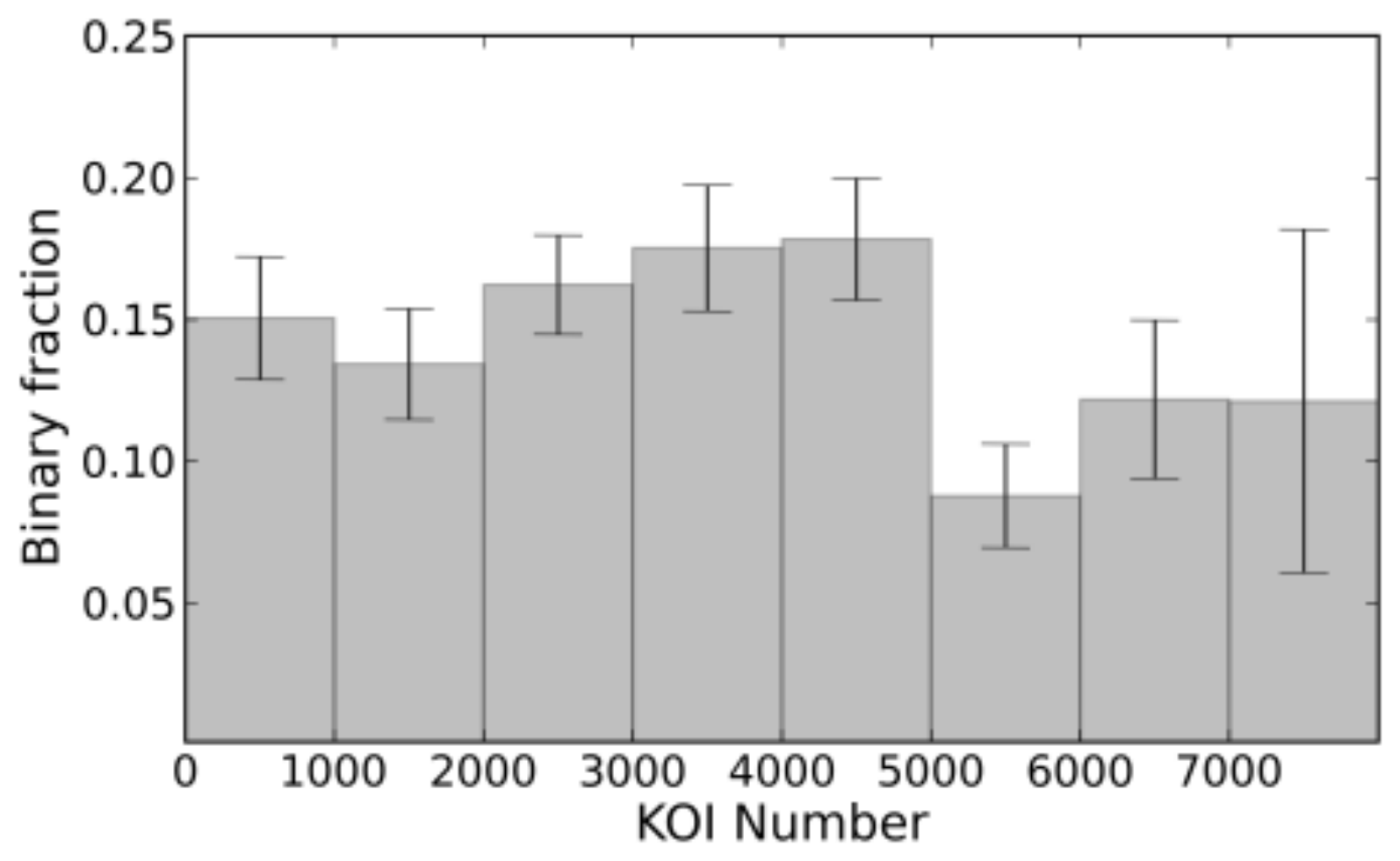}
\caption{Binary fraction within 4" of KOIs as a function of KOI number. A 2.8$\sigma$ decrease in the fraction of nearby stars between KOIs numbered less than 5000 and greater than 5000 is apparent.}
\label{fig:koinumberbinaryfrac}
\end{figure}

The initial vetting of the planetary candidates by the \textit{Kepler} team could conceivably have a built-in bias, either astrophysical in origin or as a result of their data reduction techniques, and that bias may have varied over the early and late public releases of KOIs \cite{borucki11, batalha13, burke14, coughlin15}.  A change in multiplicity with respect to KOI number may come as a result of this bias.  With a target list widely dispersed in the full KOI dataset, we can search for such a trend.  The fraction of KOIs with companions as a function of KOI number, as displayed in Figure$~\ref{fig:koinumberbinaryfrac}$, shows a sharp decrease at approximately KOI-5000.  We find KOI numbers less than 5000 have a nearby star fraction of 16.1\%$\pm$0.9\% and KOI numbers greater than 5000 have a nearby star fraction of 10.2\%$\pm$1.5\%, a 2.9$\sigma$ disparity.  The exact mechanism for this is unclear, however this may be a result of better false positive detection in the later data releases due to automation of the vetting process \cite{coughlin15}.  There is no significant corresponding variation in the separations or contrasts of stellar companions between the two populations.

\subsubsection{Stellar Multiplicity and Multiple-planet Systems}

Perturbations from the companion star will change the mutual inclinations of planets in the same system \cite{wang14}, therefore a lower number of multiple transiting planet systems are expected to have stellar companions.  Multiple planet systems are also subject to planet-planet effects \cite{rasio96, wang15a}.

\begin{figure}
\centering
\includegraphics[width=285pt]{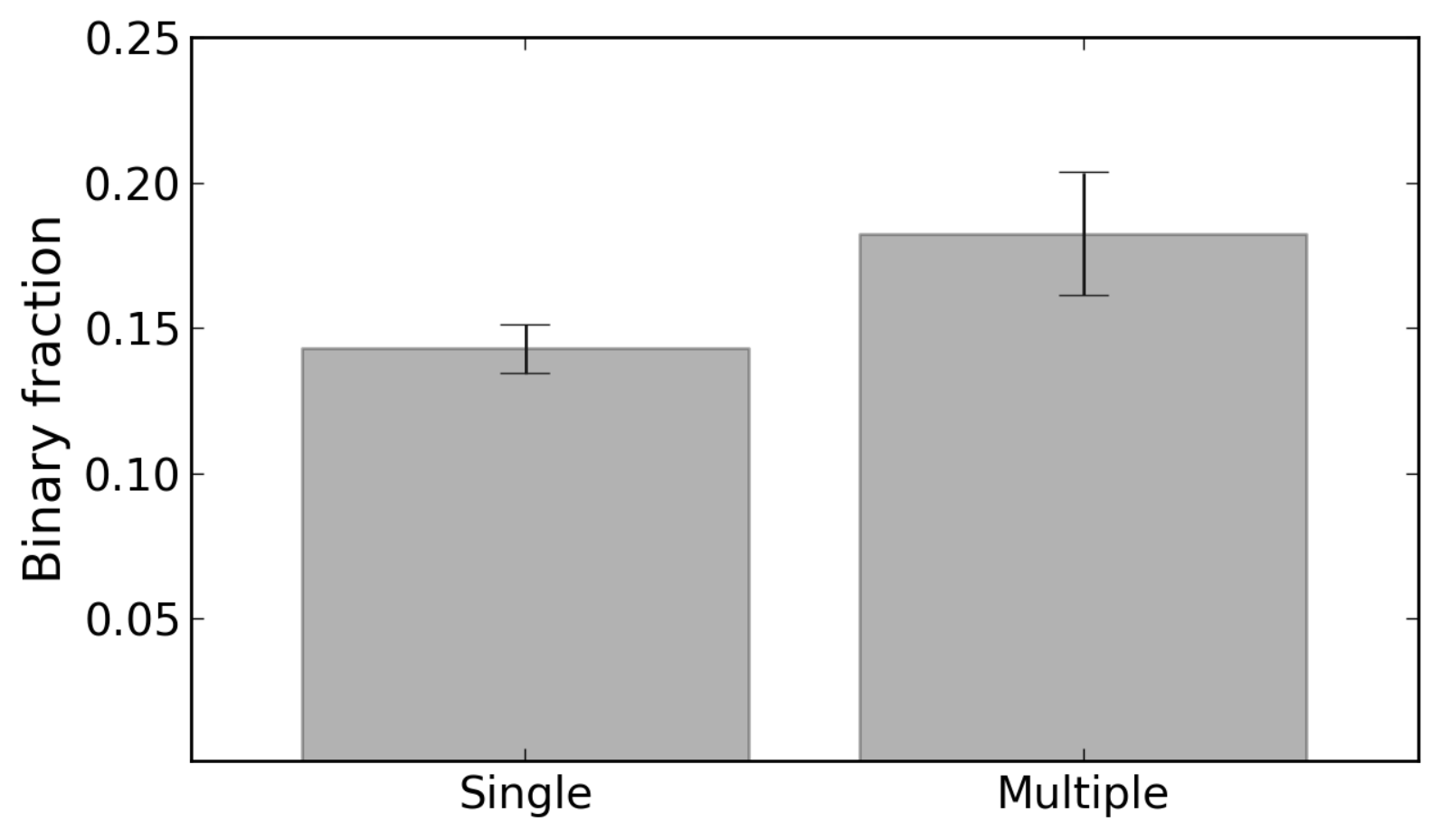}
\caption{The binary fraction within 4" of KOIs hosting detected single- and multiple-planetary systems.}
\label{fig:numplanetsbinaryfrac}
\end{figure}

In Paper I, we found a low-sigma disparity in multiplicity rates between single- and multiple- planet systems, with single-planet systems exhibiting a slightly higher binary fraction.  With a combined sample from Paper II and Paper III, we revisit this result with over three times more targets.  We find a slightly higher binary fraction for multiple planetary systems, displayed in Figure$~\ref{fig:numplanetsbinaryfrac}$.  A Fisher exact test gives a 8.7\% probability of this being a chance difference.

\subsubsection{Stellar Multiplicity and Close-in Planets}

The presence of stellar companions is hypothesized to shape the formation and evolution of planetary systems.  Overall, there is evidence that planetary formation is disrupted in close binary systems \cite{fragner11, roell12}.  The third body in the system can lead to Kozai oscillations causing orbital migration of the planets \cite{fabrycky07, katz11, naox12} or tilt the circumstellar disk \cite{batygin12}.  Smaller planets are also more prone to the influence of a stellar companion because of weaker planet-planet dynamical coupling \cite{wang15a}.  These dynamical interactions between small and large planets in the same system tend to differentially eject small planets more frequently than large planets \cite{xie14}. The presence of a stellar companion increases the frequency of these interactions, leading to higher loss of small planets.  Consequently, we would expect a correlation between binarity and planetary period for different sized planets.

\begin{figure}
\centering
\includegraphics[width=285pt]{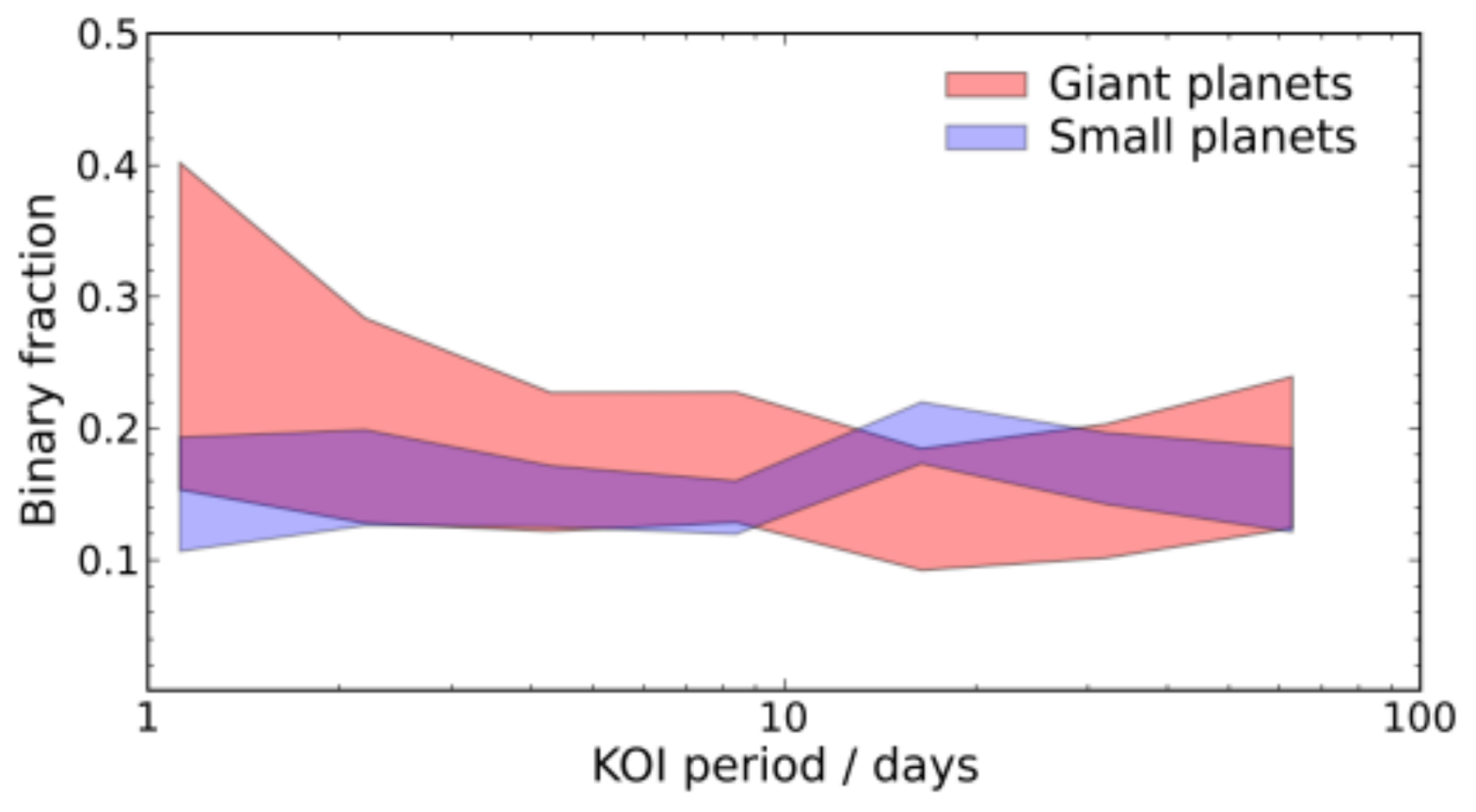}
\caption{1$\sigma$ uncertainty regions for the binary fraction as a function of KOI period for two different planetary populations.}
\label{fig:periodfrac}
\end{figure}

In Figure$~\ref{fig:periodfrac}$ the fraction of \textit{Kepler} planet candidates with nearby stars is shown, with planets grouped into two different size ranges (this analysis splits the ``small'' and ``giant'' planets at the arbitrary value of Neptune's radius, 3.9 R$_\oplus$). We again see a small increase in the nearby star fraction for giants with periods $<$15 days.

\begin{figure}
\centering
\includegraphics[width=285pt]{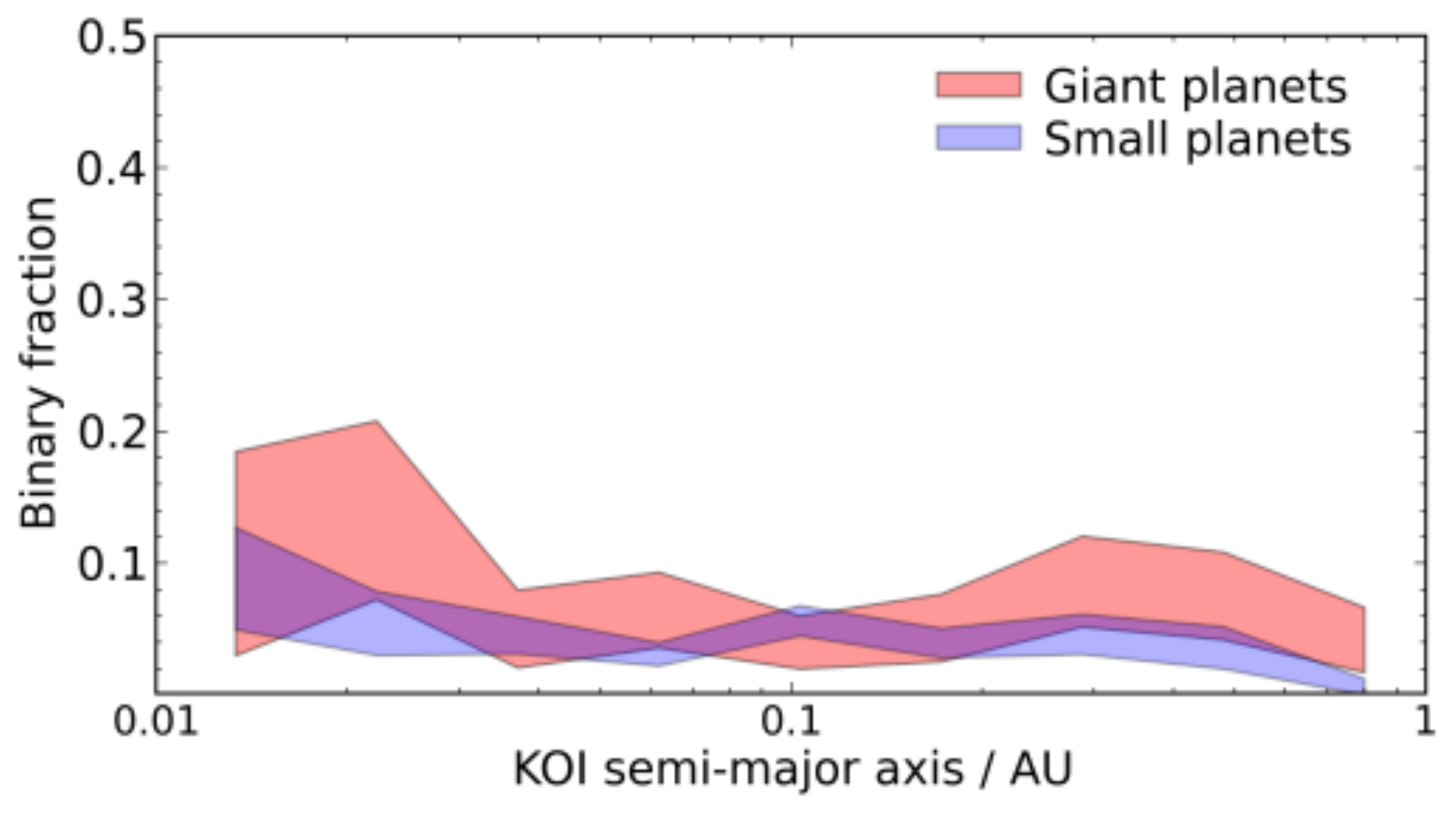}
\caption{1$\sigma$ uncertainty regions for the binary fraction as a function of KOI semi-major axis between 0.01 and 1.0 AU for two different planetary populations.}
\label{fig:semimajoraxis}
\end{figure}

Any real disparity between the populations would also manifest in the physical orbital semi-major axis, which is related to the observable periods by the stellar mass. In Figure$~\ref{fig:semimajoraxis}$ we plot the two population's binarity fraction as a function of the calculated semi-major axis of the planetary candidates between 0.01 and 1.0 AU.  No significant giant planet binarity spike is observed as in the periods plot.

This survey suggests that the presence of a second stellar body in planetary systems does not appreciably affect the number of close-in giant planets.  This agrees with the analysis of a previous study\cite{wang15a} that finds a relatively uniform multiplicity rate for planets with short and long periods.

Another previous study\cite{kraus16} found a 6.6$\sigma$ deficit in binary stars with separation $\rho$$<$50 AU in KOIs compared to field stars, again suggesting that close-in stellar companions disrupt the formation and/or evolution of planets, as had been previously hypothesized \cite{wang14}.  Indeed, a quarter of all solar-type stars in the Milky Way are disallowed from hosting planetary systems due to the influence of binary companions.

Some evidence remains, however, that stellar binarity may encourage the presence of hot Jupiters.  A recent NIR survey \cite{ngo15} of exoplanetary systems with known close-in giants finds that hot Jupiters are twice as likely as field stars to be found in a multiple star system, with a significance of 2.8$\sigma$.  However, the binarity rates of systems containing hot Jupiters remains unclear: 12$\%$ \cite{roell12}, 38$\%$ \cite{evans16}, 51$\%$ \cite{ngo15}. 

\subsubsection{Stellar Multiplicity of Planetary and Non-planetary Candidate Hosts}

The dynamical interactions between stars in binary systems create a complex environment for planet formation and evolution. The Kepler mission offers an opportunity to compare binary stars with, and without, detections of inner transiting exoplanets. Kepler stars with no planet detections can act as a control sample in an effort to discover if inner planet formation in binary systems is a rare occurrence. We build a control sample of $\sim$700 serendipitously observed targets from Robo-AO images of KOIs, while also targeting a specific control dataset with parameters matched to an initial dataset of $\sim$300 KOIs. We find that the binary fraction of KOIs, and the control sample with no detected planets, match to within 1-$\sigma$. Our findings do not suggest that inner transiting exoplanets are rare in binary systems for this sample.

\section{CONCLUSION}
\label{sec:conclusion}

We find a multiplicity rate around 3313 \textit{Kepler} planetary candidates of 14.5$\pm$0.8\% out to a separation of 4".  Within this large dataset, we find 23 potential rocky, habitable zone planetary candidates in multiple systems, 32 planetary candidates in possible multiple star systems, including 5 candidate quadruple systems hosting planets, a significant decrease in binarity between early and late \textit{Kepler} data releases, and a slightly higher binary fraction for multiple planetary systems.  We also detect a low-significance uptick in multiplicity for large, close-in planets, however this increase is not as large as theories of planetary evolution in multiple systems suggest.

The Robo-AO system was installed on the 2.1-m telescope at Kitt Peak in November 2015, and a new low-noise infrared camera that will allow observations of redder companion stars will be added in the future\cite{salama16}.  In addition, a second generation Robo-AO instrument on the University of Hawai`i 2.2-m telescope on Maunakea \cite{Robo-AO2, baranecspie16} is being built.  The two systems will together image up to $\sim$500 objects per night and have access to three-quarters of the sky over the course of a year.  A southern analog to Robo-AO, mounted on the Southern Astrophysical Research Telescope (SOAR) at CTIO and capable of twice HST resolution imaging, is also in development.  With unmatched efficiency, Robo-AO and its lineage of instruments are uniquely able to perform high-acuity imaging of the hundreds of K2 \cite{K2} planetary candidates, ground-based transit surveys such as MEarth \cite{mearth}, KELT \cite{kelt1, kelt2}, HATNet \cite{hatnet}, SuperWASP \cite{superwasp}, NGTS \cite{ngts}, XO \cite{xo}, and the Evryscope \cite{evryscope}, as well as the thousands of expected exoplanet hosts discovered by the forthcoming NASA Transiting Exoplanet Survey Satellite \cite{TESS} and ESA PLAnetary Transits and Oscillations of stars 2.0 \cite{PLATO} missions.  

\acknowledgments
 
This research is supported by the NASA Exoplanets Research Program, grant $\#$NNX 15AC91G. C.B. acknowledges support from the Alfred P. Sloan Foundation. T.M is supported by NASA grant $\#$NNX 14AE11G under the \textit{Kepler} Participating Scientist Program. 

The Robo-AO system is supported by collaborating partner institutions, the California Institute of Technology and the Inter-University Centre for Astronomy and Astrophysics, and by the National Science Foundation under Grant Nos. AST-0906060, AST-0960343, and AST-1207891, by the Mount Cuba Astronomical Foundation, and by a gift from Samuel Oschin. We are grateful to the Palomar Observatory staff for their ongoing support of Robo-AO on the 1.5-m telescope, particularly S. Kunsman, M. Doyle, J. Henning, R. Walters, G. Van Idsinga, B. Baker, K. Dunscombe and D. Roderick.

\bibliography{main} 
\bibliographystyle{spiebib} 

\end{document}